\documentclass[fleqn,usenatbib]{mnras}

\usepackage{newtxtext,newtxmath}
\usepackage{orcidlink}

\usepackage[T1]{fontenc}

\DeclareRobustCommand{\VAN}[3]{#2}
\let\VANthebibliography\thebibliography
\def\thebibliography{\DeclareRobustCommand{\VAN}[3]{##3}\VANthebibliography}
\newcommand{\add}[1]{\textcolor{black}{#1}}
\newcommand{\addr}[1]{\textcolor{black}{#1}}

\usepackage{graphicx}	% Including figure files
\usepackage{amsmath}	% Advanced maths commands
\title[Bar structure and disk bending wave at z = 4.4]{Detecting a disk bending wave in a barred-spiral galaxy at redshift 4.4}

\author[T. Tsukui et al.]{
Takafumi Tsukui,$^{\orcidlink{0000-0002-1499-6377},1,2,3,4}$\thanks{E-mail: tsukuitk23@gmail.com (TT)}
Emily Wisnioski,$^{\orcidlink{0000-0003-1657-7878},1,2}$
Joss Bland-Hawthorn,$^{\orcidlink{0000-0001-7516-4016},2,5}$ 
Yifan Mai,$^{\orcidlink{0000-0003-3514-6280},2,5}$
Satoru Iguchi,$^{\orcidlink{0000-0002-7191-2301},3,4}$
\newauthor
\ \ Junichi Baba,$^{\orcidlink{0000-0002-2154-8740},6,3,4}$
Ken Freeman$^{\orcidlink{0000-0001-6280-1207},1,2}$\\
$^{1}$Research School of Astronomy and Astrophysics, Australian National University, Cotter Road, Weston Creek, ACT 2611, Australia\\
$^{2}$ARC Centre of Excellence for All Sky Astrophysics in 3 Dimensions (ASTRO 3D), Australia\\
$^{3}$ National Astronomical Observatory of Japan, National Institute of Natural Sciences, 2-21-1 Osawa, Mitaka, Tokyo, Japan \\
$^{4}$ Department of Astronomical Science, SOKENDAI (The Graduate University for Advanced
Studies) 2-21-1 Osawa, Mitaka, Tokyo, Japan\\
$^{5}$ Sydney Institute for Astronomy, School of Physics, A28, The University of Sydney, NSW 2006, Australia\\
$^{6}$ Amanogawa Galaxy Astronomy Research Center, Kagoshima University, 1-21-35 Korimoto, Kagoshima 890-0065, Japan
}

\date{Accepted XXX. Received YYY; in original form ZZZ}

\pubyear{2023}

\begin{document}
\label{firstpage}
\pagerange{\pageref{firstpage}--\pageref{lastpage}}
\maketitle

\begin{abstract}
The recent discovery of barred spiral galaxies in the early universe ($z>2$) poses questions of how these structures form and how they influence galaxy \add{evolution} in the early universe. In this study, we investigate the morphology and kinematics of the far infrared (FIR) continuum and [C\textsc{ii}] emission in BRI1335-0417 at $z\approx 4.4$ from ALMA observations. %The variations in position angle and ellipticity of the broadband isophotes as a function of deprojected radius are indicative of a barred galaxy.
The variations in position angle and ellipticity of the isophotes show the characteristic signature of a barred galaxy. The bar, $3.3^{+0.2}_{-0.2}$ kpc long in radius and bridging the previously identified two-armed spiral, is evident in both [C\textsc{ii}] and FIR images, driving the galaxy's rapid evolution by channelling gas towards the nucleus.
%explain indicating its role in driving the rapid galaxy evolution by hosting star formation and channelling gas towards the centre and is potentially responsible for the bright quasar. %Fourier series expansion of the azimuthal velocity profile along the concentric ellipses 
Fourier analysis of the [C\textsc{ii}] velocity field reveals an unambiguous \add{kinematic} $m=2$ mode with a line-of-sight velocity amplitude of up to $\sim30-40$ km s$^{-1}$; a plausible explanation is the disk's vertical bending mode triggered by external perturbation, which presumably induced the high star formation rate and the bar/spiral structure.
%JBH: use 70 rather than 73 when a limit is used, not precise estimation
The bar identified in [C\textsc{ii}] and FIR images of the gas-rich disk galaxy ($\gtrsim 70$\% of the total mass within radius $R\approx 2.2$ disk scale lengths) suggests a new perspective of early bar formation \add{in high redshift gas-rich galaxies} -- a gravitationally unstable gas-rich disk creating a star-forming gaseous bar, rather than a stellar bar emerging from a pre-existing stellar disk. \add{This may explain the prevalent bar-like structures seen in FIR images of high-redshift submillimeter galaxies.}
%JBH: this last sentence below could be dropped since you already have speculation above. if you need to cut lines
%This study lends support to the use of disk corrugations as evidence of a recent strong perturbation.
%This study observationally demonstrates a seismic ripple in the disk can be detectable and used as evidence of a recent interaction.
\end{abstract}

\begin{keywords}
galaxies: bar -- galaxies: disc -- galaxies: kinematics and dynamics -- galaxies: high-redshift -- galaxies: spiral
\end{keywords}

%%%%%%%%%%%%%%%%%%%%%%%%%%%%%%%%%%%%%%%%%%%%%%%%%%

%%%%%%%%%%%%%%%%% BODY OF PAPER %%%%%%%%%%%%%%%%%%

\section{Introduction}
Bar structure plays a crucial role in driving galaxy evolution and shaping disk structure. In galaxies an axisymmetric \add{stellar} bar exerts gravitational torque on the gas, driving it towards the galactic centre and forming a centralized stellar structure such as a bulge and nuclear disk \citep{Athanassoula1992-rc, Wada1992-xc}. This process may also promote gas accretion onto the black hole observed as active galactic nuclei \citep[AGN;][]{Emsellem2015-kx, Hopkins2010-bo}. Bars can also drive radial migration of gas and stars, which is essential for explaining the observed stellar kinematics in Milky Way galaxies \citep[e.g.][]{Kawata2021-mv}. %Also, the stellar bar structure is susceptible to buckling instability thickening stellar disk \citep[e.g.][]{Raha1991-im}.

Numerical simulation suggests that \add{stellar} bar formation in galaxies leads to an immediate gas inflow into the central region and the formation of a nuclear disk, making the stellar age of the nuclear disk a good indicator of the epoch when the bar first formed \citep{Baba2020-ld}. \add{Rather earlier however, the idea was applied to observations for estimating bar formation age in \citet{Gadotti2015-pu}.}

%JBH: IFU is one tiny button, Hector has 20 IFUs working in an IFS. Lots get this wrong!
Recent observations using integral field spectroscopy (IFS) have provided insight into the stellar populations of nuclear disks in barred galaxies \citep{Gadotti2015-pu, Bittner2020-qs, deSaFreitas_2023-kf}, suggesting that the oldest nuclear disks in barred galaxies are at least 10 Gyr old ($z>2$; \citealt[][]{De_Sa-Freitas2023-rd}). The findings, pointing to early bar formations, align with the latest high redshift observations. For instance, recent Atacama Large Millimeter/submillimeter Array (ALMA) observations show gas disks already formed at $z>4$ \citep[e.g.,][]{Neeleman2020-zu,Rizzo2020-yc, Lelli2021-hh, Tsukui2021-vl}, while James Webb Space Telescope (JWST) observations recently discovered early barred galaxies at redshift $1<z<3$ \add{\citep{Guo2022-dr, Le_Conte2023-an}} and at $z=4.2$ \citep{Smail2023-gm}. In addition, numerous disk-like systems ($\sim50\%$) have been found at $3<z<6$ by JWST \citep{Ferreira2022-ss, Nelson2023-cg}. \add{While t}hese recent discoveries are consistent with some pre-JWST and ALMA studies showing disk prevalence already at $z\sim2.6$ \citep[][but see \citealt{Conselice2014-rz}]{Wuyts2011-iw, Wisnioski2015-ah}\add{,} they show surprisingly numerous disks up to $z\sim6$ and earlier spiral and bars than expected \citep{Elmegreen2014-xx}.

%These recent discoveries contrast with earlier claims that disks only begin to form at redshifts $\lesssim 2$ established with pre-JWST and ALMA observations \citep[e.g.,][]{Conselice2014-rz}.

Numerical simulations motivated by these discoveries show that baryon-dominated disk galaxies promptly form a bar \add{\citep{Fujii2018-ji, Bland-Hawthorn2023-yc}}. It remains an open question if this rule applies to galaxies with high gas fractions \citep[more than $\sim$ 50\% of total baryonic mass at $z>3$;][]{Carilli2013-im}, as some simulations suggest molecular gas can suppress bar formation or result in a weaker bar \citep{Lokas2020-nz, Athanassoula2013-st}. \add{Conversely, ALMA observations reveal prevalent bar morphologies in dust continuum images of gas-rich submillimeter galaxies \citep{Gullberg2019-ec, Hodge2019-uo, Smail2023-gm}.} 
Tidal interaction is another promising avenue to form bars \citep[e.g.,][]{Noguchi1996-fo, Lokas2014-pd, Lokas2016-nx}, even for gas-rich systems \citep{Gajda2018-jq}. Recent cosmological simulations, including a realistic high gas fraction of high-redshift galaxies and external effects such as mergers, suggest a high bar fraction out to $z\sim4$ \citep{Rosas-Guevara2022-qx}. \add{Different sub-grid physics implementations in simulations can result in different galaxy properties and bar dynamics \citep[][]{Fragkoudi2021-vn}}.

%, although the resolution may not necessarily be optimized to resolve the 3d disk perturbation essential for bar formation for low mass system \citep{Wilkinson2023-ag}. 

%Despite the potential for bars to form in a high redshift universe, the observational bias is significant for identifying barred galaxy populations including observational sensitivity, resolution, and wavebands \citep{Erwin2018-xh, Rosas-Guevara2022-qx}. As barred galaxies may on average experience bursty star formation \citep{Carles2016-fu, Fraser-McKelvie2020-mp}, the earliest bar formation may be heavily dust-obscured, and some may be elusive with Hubble space telescope (HST) \citep{Smail2023-gm} or even the modest depth JWST observation \citep{Kokorev2023-mc}. 

%ALMA observation at $860\micron$ band shows rich structure in dusty starforming galaxies including bar-like axisymmetric structures at redshift z=1.5 to 4.8 within the disk morphologies \citep{Hodge2019-uo}. Given the ALMA's exquisite sensitivity and spatial resolution, $860\micron$ observation may be a promising window to reveal the earliest population of barred galaxies and the bar-driven galaxy evolution, providing dust extinction-free star formation or young stellar population tracer, and an almost constant flux as a function of redshift for the same star formation rate \citep[negative k-corrections;][]{Guiderdoni1997-ax}. 

Recently, \citealt{Tsukui2021-vl} revealed a spiral morphology in BRI 1335-0417 at $z=4.4074$ \citep{Guilloteau1997-xe}, hosting an optical quasar initially identified in Automatic Plate Measuring (APM) survey \citep{Storrie-Lombardi1996-rw}. The galaxy exhibits a high star formation rate (SFR) $\sim 1700 M_{\odot}$~yr$^{-1}$ estimated from the spectral energy distribution (SED) modelling with the AGN contribution being corrected using spatially resolved information (point source and extended source separation; \citealt{Tsukui2023-wo}), making it one of the brightest unlensed submillimeter source at $z>4$ \citep{Jones2016-cj}. The [C\textsc{ii}] and dust continuum maps of the galaxy revealed a two-armed structure with a pitch angle of ${26.7^{+4.1}_{-1.6}}^\circ$ \citep{Tsukui2021-vl}. These arms extend from 2~kpc to 5~kpc and appear to start at the end of an elongated bar-like structure that bridges them. 

Despite new observational and theoretical results, it remains unclear what was the dominant cause for early bar and spiral formation -- internal or external processes. As the brightest and earliest barred spiral example, BRI 1335-0417 allows us to study the detailed [C\textsc{ii}] line and FIR continuum morphology in unprecedented detail and without uncertainties due to lens model reconstruction. The spatially resolved [C\textsc{ii}] line kinematics together with numerical simulations provide an excellent laboratory to provide new insights into early bar formation. 

 %$158\micron$ ionized carbon [C\textsc{ii}] line is the main coolant of the interstellar medium and thus one of the most brightest lines in FIR bands. For galaxies at redshift $z\sim4$, $158\micron$ ionized carbon [C\textsc{ii}] line is redshifted into $860\micron$ band and used as a powerful tracer of the gas disk dynamics, determining the dynamical states and thus the origin of the spiral/.

Gas disk kinematics tell us not only the dynamics of the galaxy, such as disk stability and underlying mass distribution of the galaxy. Subtracting the overall rotation and examining the more subtle residual velocity field allows us to explore further: gas inflows \citep[e.g.,][]{Di_Teodoro2021-ll, Genzel2023-lj}, bar and dynamical effect of the bar/spiral structure \citep[e.g.,][]{Grand2016-ys, Gomez2021-yx, Monari2016-ro}, and even bending waves of the disk --- seismic ripple propagating through the disk due to perturbation by a recent interaction with satellites \citep{Bland-Hawthorn2021-fh, Tepper-Garcia2022-sh, Urrejola-Mora2022-in} \add{or misaligned gas accretion \citep{Khachaturyants2022-ip}}.

In this \add{paper}, we report a new analysis of the bar structure in the quasar host galaxy BRI 1335-0417 using ALMA Band 7 data of the far infrared (FIR) continuum (observed frame $\sim 869\micron$ or rest-frame $\sim \add{160}\micron$) and ionized carbon [C\textsc{ii}]. We also demonstrate that the rotation-subtracted residual velocity field is consistent with
%suggests the characteristic signature of 
the dynamical imprint of a recent interaction by an external perturber, which likely induced the bar and spiral density wave in the gas disk. 
% don't conjecture yet - too soon

Throughout this paper, we assume a flat $\Lambda$-dominated cold dark matter ($\Lambda$CDM) cosmology with a present-day Hubble constant of $H_0=70$ km s$^{-1}$ Mpc$^{-1}$, and a density parameter of pressureless matter $\Omega_M=0.3$.

\section{Observation and data reduction}
This study uses \add{rest-frame $160\micron$} FIR continuum image and [C\textsc{ii}] line cube from the observation program \#2017.1.00394.S (PI=González López, Jorge), which was carried out on 2018 January 21. The data calibration and reduction details were described in \citep{Tsukui2021-vl, Tsukui2023-wo}. In order to accurately quantify the elongated bar structure, both [C\textsc{ii}] and FIR intensity maps were convolved to have a circular-shaped beam (point spread function; PSF) with the same resolution (Full-width half maximum (FWHM) = $0.195$ arcsec = $1.3$ kpc). \add{The [C\textsc{ii}] line data are binned in the spectral axis to have a spectral resolution of 20 km s$^{-1}$.}

\begin{figure*}
	\includegraphics[width=0.99\textwidth]{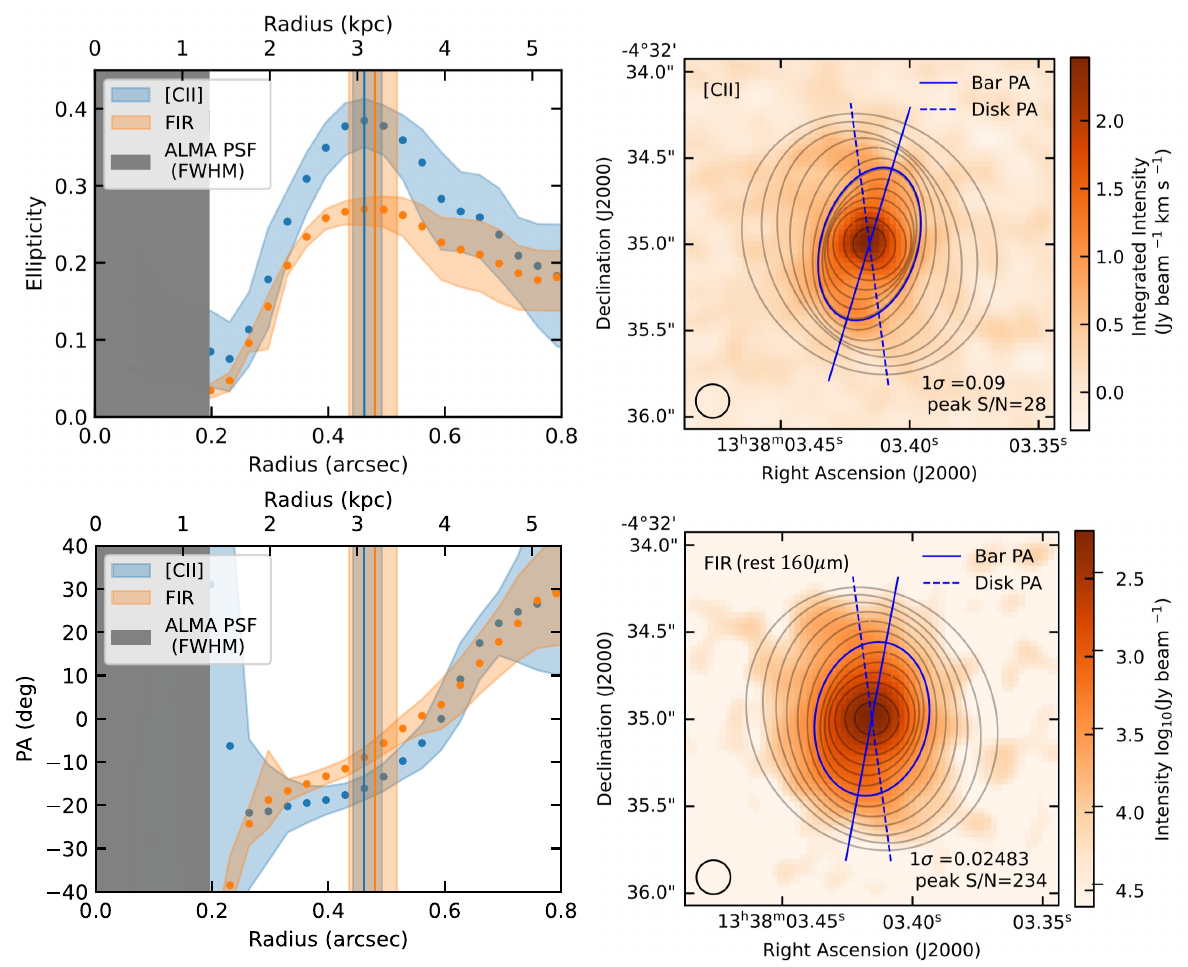}
    \caption{Left column: ellipticity (top) and position angle (PA; bottom) of isophote ellipses as a function of the ellipse radius (semi-major axis) of [C\textsc{ii}] (blue) and FIR (orange) images of BRI 1335-0417. The vertical lines indicate the radii of the maximum ellipticity for [C\textsc{ii}] (blue) and FIR (orange). For each measurement, 1$\sigma$ uncertainty is shown with a shaded region. The grey-shaded region indicates the PSF FWHM, where accurate ellipticity and PA estimates are not possible. Right column: [C\textsc{ii}] and FIR images of BRI 1335-0417 overlaid with the best-fitting isophote ellipses in black, the bar ellipse and its PA in the solid blue line. The disk kinematic PA \citep{Tsukui2021-vl} is also shown in the dashed blue line. The FWHM of PSF is shown in the bottom left corner of both images. }
    \label{fig:Fig1}
\end{figure*}

\section{Results and Discussion}
In this section, we present the results of structural investigations in the first two subsections and then investigations from the dynamical point of view using first-order and higher-order [C\textsc{ii}] kinematics in the subsequent subsections.
\subsection{Bar identification by the Ellipse fitting}

% JBH: use present tense, much better to read, more dynamic
We employ the commonly used ellipse-fitting method to examine the photometric structure of the disk \add{\citep[e.g.,][]{Jedrzejewski1987-js, Wozniak1995-su, Erwin2005-lf, Gadotti2007-gy}}. 
%JBH: don't say confirm because it sounds like you know the answer < you are right.
We fit ellipses to isophotes of the [C\textsc{ii}] and FIR intensity images, with position angle (PA) and ellipticity ($\epsilon$) allowed to vary for each ellipse and the central position fixed to the best fit of the smallest ellipse. To estimate the statistical uncertainty in our measurements, we performed the fitting procedure 300 times, each time adding realistic correlated noise to the original images. The noise properties are measured using the noise auto-correlation function \citep{Tsukui2023-fv}. 

% overlain is too poetic
The left column of Fig.~\ref{fig:Fig1} shows the ellipticity and position angle of the best-fit ellipses as a function of radius. The right column presents the best-fit isophote ellipses overlaid on [C\textsc{ii}] and FIR images. The ellipticity profile of the best-fit ellipses exhibits a characteristic profile common to barred galaxies \citep{Wozniak1995-su}. At smaller radii, the ellipticity is small as the ellipses trace the centrally concentrated light. Then the ellipticity reaches a maximum as the ellipse aligns with the elongated bar shape. Subsequently, the ellipticity decreases as the ellipse traces the disk more circular than the bar. The position angle (PA) changes slowly as the ellipse traces the bar and rapidly changes at the end of the bar as the ellipse starts tracing the disk because the disk position angle is offset from the bar in most cases\footnote{The elongated bar structure has a random orientation relative to the disk position angle.}. 

In both [C\textsc{ii}] and FIR continuum, the ellipticity reaches a maximum at the same radius within some uncertainty. Therefore, we adopt the radius of the maximum ellipticity of [C\textsc{ii}] as a fiducial sky-projected bar length, $R^{\mathrm{sky}}_{\textrm{bar}, \epsilon_\textrm{max}}=3.1^{+0.2}_{-0.1}$ kpc. This bar length is larger than the 2$\times$FWHM of the beam, which is the required minimum criterion for detecting the bar \citep{Erwin2018-xh}. Another way to define the bar length from the ellipse fitting is the radius where the position angle (PA) changes by 5 degrees relative to its value at the radius of maximum ellipticity. This alternative definition provides slightly larger values, $R^{\mathrm{sky}}_{\textrm{bar}, \Delta\textrm{PA}=5\textrm{deg}}=3.5^{+0.3}_{-0.1}$ kpc yet similar to our adopted fiducial value, confirming that the choice of the definition does not affect the conclusion.

We find the intrinsic bar length $R^{\mathrm{int}}_{\textrm{bar}, \epsilon_\textrm{max}}=3.3^{+0.2}_{-0.2}$ kpc by the analytical deprojection assuming a planer bar ellipse \citep{Gadotti2007-gy} and the disk inclination \add{(37.3 deg)} and position angle \add{(7.6 deg)} estimated by \citet{Tsukui2021-vl}. The bar length relative to the [C\textsc{ii}] disk scale length $R_{\mathrm{d}}=1.83\pm0.04$ kpc \citep{Tsukui2021-vl} is larger than stellar bars seen in nearby galaxies \citep[][their derived scaling relation predicts a bar length of $1.7\pm0.1$ kpc given the same disk scale length]{Erwin2019-im}. 

\add{We note that the PA of the outermost isophote is misaligned with the kinematic disk PA. The outer ellipses are influenced not only by noise, as evidenced by the increased uncertainty shown in the bottom left panel of Fig.\ref{fig:Fig1}, but also by the spiral structure and faint tidal tail-like structure extending from north-east to south-west. Therefore, in later analysis, we will use the kinematic disk PA rather than the photometric isophote PA as the disk's position angle.}

\subsection{Interpretation of [C\textsc{ii}] and FIR continuum bars}
The [C\textsc{ii}] line traces not only overall star formation but also multiphase gas distribution, from neutral gas \citep{Herrera-Camus2018-vf} to molecular and ionized gas \citep{Pineda2013-wx}. On the other hand, the FIR continuum represents thermal emission from dust heated by ultraviolet emission produced by young massive stars. Although the FIR continuum is commonly used as the tracer for dust mass and star formation rate \add{(young massive stars heating the dust)}, the relationships of the single band FIR continuum to these two quantities are complex depending on the spatial temperature and opacity variation \citep{Da_Cunha2021-ma, Tsukui2023-wo}. The FIR continuum is proportional to the dust mass and star formation rate\footnote{With several assumptions such that the dust is solely heated by the young massive stars and archived single equilibrium temperature. Stellar initial mass function (IFM) and dust geometry are known.} specifically in a limiting case of dust being optically thin with a spatially constant temperature. 

\begin{figure}
	\includegraphics[width=0.99\columnwidth]{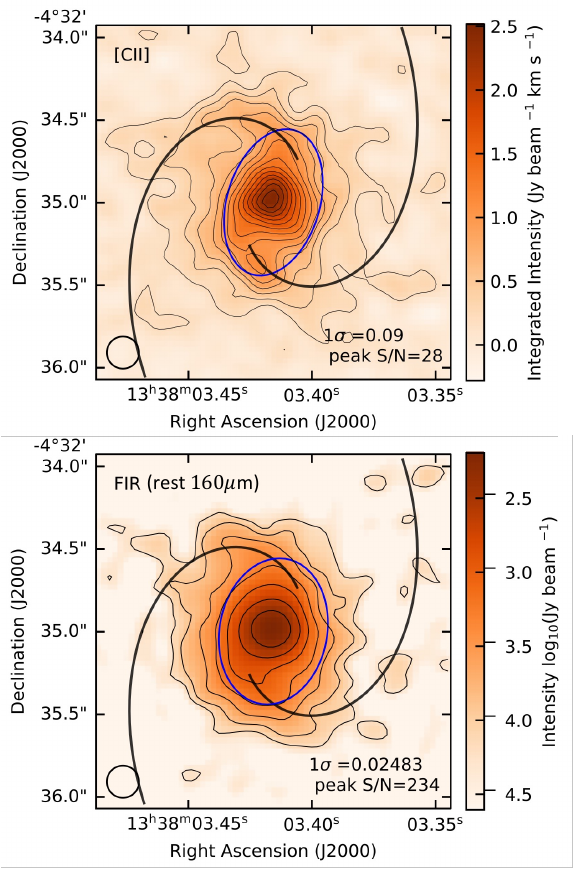}
    \caption{[C\textsc{ii}] and FIR images of BRI 1335-0417 overlaid with our identified bar ellipses (blue line) and two-armed spiral structure identified in \citet[black solid line]{Tsukui2021-vl}. Contours start at 2$\sigma$ in both maps but are linearly spaced by 2$\sigma$ for [C\textsc{ii}] map and logarithmically spaced in powers of 2 for FIR map ($2\sigma, 4\sigma$, $8\sigma$, ...).}
    \label{fig:Fig2}
\end{figure}

The ellipse fitting result confirms both [C\textsc{ii}] and FIR continuum images have elliptical elongated bar shapes, which are shown in Fig.~\ref{fig:Fig2} along with the two-armed spiral confirmed in \citealt{Tsukui2021-vl} (projected on the sky). The configuration of the bar and spiral is similar to the nearby grand-design spiral galaxies observed \citep[e.g.,][]{Stuber2023-vk} and found in simulations \citep[e.g.,][]{Baba2015-vr}. 

In the nearby universe, not all barred galaxies show star formation and gas reservoirs in the bar region. Star formation or gas lanes along the bar region are preferentially found in gas-rich late-type barred galaxies \citep{Diaz-Garcia2020-le, Fraser-McKelvie2020-mp} and more massive gas-rich barred systems \citep{Stuber2023-vk}. FIR continuum bars are often seen in high redshift ($z\sim 2$ to 4) submillimeter galaxies \citep{Hodge2019-uo, Gullberg2019-ec}, implying that the SF and gas bar may be common in the gas-rich early galaxies \citep{Carilli2013-im}. 

%showed the visible gas lane aligned with the stellar bar is found preferentially in the more massive and gas-rich system in the main sequence nearby barred galaxies. 

Interestingly, there is a slight yet statistically significant offset in PA of the bar ellipses between [C\textsc{ii}] and FIR continuum (PA$_{[C\textsc{ii}]}=-15.9^{+4.0}_{-3.1}$, PA$_{\mathrm{FIR}}=-6.9^{+4.2}_{-4.7}$). Also, the bar ellipse of [C\textsc{ii}] is more elongated than that of the FIR continuum ($\epsilon_{[C\textsc{ii}]}=0.27^{+0.02}_{-0.02}$, $\epsilon_{\mathrm{FIR}}=0.39^{+0.03}_{-0.02}$). Although the rounder shape of the FIR bar may be attributed to the significant AGN contribution to the FIR image suggested by \citet{Tsukui2023-wo} these morphological differences, if further confirmed by higher resolution data, may tell the dynamics of the bar.

These differences may be consistent with the bar characteristics commonly confirmed in local observations \citep{Fraser-McKelvie2020-mp} and simulation \citep{Emsellem2015-kx} -- the compressed gas flow (traced by [C\textsc{ii}]) which usually forms a straight line into the centre of the galaxy along the leading edge of the bar made up by the stellar components (traced by FIR emission), making the [C\textsc{ii}] bar thinner than the FIR bar. The [C\textsc{ii}] bar is also inclined towards the leading edge of the FIR bar. 

This interpretation requires two assumptions. (1) The spiral arm in this galaxy is trailing as the majority of spiral galaxies are trailing \citep[e.g.,][]{Iye2019-nt}. If so, the right (west) side in Fig.~\ref{fig:Fig2} is the far side of the disk, and the disk rotates clockwise, given that the north/south is the receding/approaching side of the disk (see Fig.~\ref{fig:Fig3}A). The bar ellipse of the [C\textsc{ii}] line is displaced towards the leading edge of the bar (on the downstream side of the gas motion). (2) \add{The FIR emission may trace more towards young stellar components than interstellar medium (ISM) traced by [C\textsc{ii}], as the FIR emission can depend on the distribution of the dust (ISM) and young massive stars which heat the dust}. Especially in the early universe, $\sim$ 1~Gyr, the main producers of dust grains are asymptotic-giant-branch (AGB) stars or core-collapse supernovae \citep{Gall2011-wf}. AGB stars start contributing at 10~Myr to 100~Myr after the stellar evolution of stars with the mass of 3-8$M_{\odot}$ while core-collapse supernovae contribute after the lifetime of 10~Myr. Within the relatively longer dynamical time scale of the disk, 120~Myr, the stellar distribution producing the dust would not have displaced relative to the dust distribution. 

As discussed in the next section, the stellar content is quite small relative to gas, so the bar dynamics and formation may differ from the nearby stellar-dominated bar paradigm. 

\subsection{Dynamical constraint from [C\textsc{ii}] rotation curve}
The detailed mass distribution within galaxies is an essential factor for bar formation \add{\citep[e.g.,][]{Efstathiou1982-pq, Fujii2018-ji, Romeo2023-dw, Bland-Hawthorn2023-yc}}. A disk-dominated system is prone to instability and bar formation, while spherical structures such as dark matter and bulges as well as the presence of gas can suppress the bar formation. In this section, we derive a lower boundary on the disk fraction and gas fraction using (1) the dynamical mass distribution estimated by \citet{Tsukui2021-vl} whose rotation curve is consistent with the independent modelling by \citet{Roman-Oliveira2023-uh} and (2) the CO(2-1) line luminosity which traces molecular gas mass \citep{Jones2016-cj}. Then, we discuss the implication for the bar formation in this galaxy.

\citealt{Tsukui2021-vl} decomposed the rotation curve into contributions from the bulge and disk, assuming a de Vaucouleurs mass distribution for the bulge and an exponential for the disk (with the scale radius from the [C\textsc{ii}] emission). The derived disk mass, $M_\mathrm{disk}=4.9^{+1.7}_{-2.5}\times10^{10}M_{\odot}$, is consistent with the lower limit of the molecular gas mass, $M_\mathrm{gas}=5.1\times10^{10}M_{\odot}$ estimated by \citealt{Jones2016-cj} assuming solar metalicity and a typical CO line ratio $r_{21}=L_{\mathrm{CO(2-1)}}/L_{\mathrm{CO(1-0)}}=0.85$ for submillimeter galaxies \citep{Carilli2013-im}, suggesting that the galaxy is baryon-dominated and gas-rich. 

As \citet{Tsukui2023-wo} revealed the optical emission of BRI 1335-0417 is dominated by a quasar, using a typical $r_{21}=0.99$ for quasars \citep{Carilli2013-im} further reduces the estimated molecular gas mass by 10\% to $M_\mathrm{gas}=4.6\times10^{10}M_{\odot}$. Using this lower limit on the molecular gas estimate, we derive a lower limit for the disk mass fraction within the radius at which disk dynamics dominates, $2.2R_d = 4.0$~kpc, 
\begin{equation}
f_{\mathrm{disk}}=\left(\frac{v_{\mathrm{disk}}(R)}{v_{\mathrm{total}}(R)}\right)^2_{R=2.2R_{\mathrm{disk}}}>\left(\frac{v_{\mathrm{gas}}(R)}{v_{\mathrm{total}}(R)}\right)^2_{R=2.2R_{\mathrm{disk}}}>0.73\pm0.07
\end{equation}
 where $v_{\mathrm{total}}(2.2R_d)=200\pm10$~km~s$^{-1}$ is the total circular velocity at $2.2R_d$ \citep{Tsukui2021-vl}, and $v_{\mathrm{gas}}(2.2R_d)=179$~km~s$^{-1}$ is the circular velocity of the exponential gas disk at $2.2R_d$ with the lower limit mass \footnote{We assumed the dispersion supported disk with finite thickness and the scale radius $R_d$ of [C\textsc{ii}] \citep{Tsukui2021-vl}.}. This lower limit on the disk mass fraction can also be interpreted as a lower limit on the gas mass fraction and baryon fraction in the disk, suggesting that the gravitational potential of BRI 1335-0417 is dominated by gas rather than stars and dark halo \citep[see also][suggesting gas dominance in this galaxy]{Carilli2002-vo, Riechers2008-gh}. 

Recent numerical studies have primarily focused on forming stellar bars out of stellar disks alternating the orbits of disk stars \citep{Bland-Hawthorn2023-yc} in which the gas delay the process. However, a disk with such a high gas fraction as in BRI 1335-0417 may behave differently. The dominant gas disk could potentially lead to the formation of a gas bar \citep{Barnes2001-sj}, as opposed to a dominant stellar disk forming a stellar bar, which then influences the gas kinematics. Theoretically, an axisymmetric 100\% gas disk is found to be able to form a self-gravitating stable gaseous bar structure \citep[][]{Cazes2000-eu}. This scenario seems plausible explaining the prevalent high-redshift bar structures seen in far-infrared emissions \citep{Hodge2019-uo, Gullberg2019-ec} and another high-z barred galaxy at $z=4.3$ with an extreme gas fraction \citep{Smail2023-gm}. \add{However, whether the stars forming out of a gaseous bar can lead to a stellar bar as we see in later epochs \citep{Guo2022-dr, Le_Conte2023-an} remains an open question, as only idealised simulation experiments have been conducted thus far.}

%In the small gas fraction regime $f_{\mathrm{gas}}\sim0.1-0.3$ the stellar bar formation by deforming the orbit of existing disk stars is suppressed \citep{Lokas2020-nz, Athanassoula2013-st}. 

%With the lack of stellar images, comparing the dynamical mass and the gas mass estimate may provide useful constraints on the dark matter fraction and the relative importance of stellar and gas potential. \citealt{Tsukui2021-vl} derived the dynamical mass distribution in BRI 1335-0417 using the first-order rotation curve and correcting the pressure support of the gas dispersion. Here, we summarize the baryon dominance and mass allowance for the stars in BRI 1335-0417 and discuss the implication along with the bar identification in the previous section. 

%We first derive the lower limit of the disk mass fraction within the radius 2.2$\times R_d$. 

%Comparing the derived dynamical mass with the lower limit of the molecular gas mass derived by \citep{Jones2016-cj} provides the upper limit of dark matter or the lower limit of the gas mass fraction. 

\subsection{2nd order disk kinematics}
\begin{figure*}
	\includegraphics[width=0.99\textwidth]{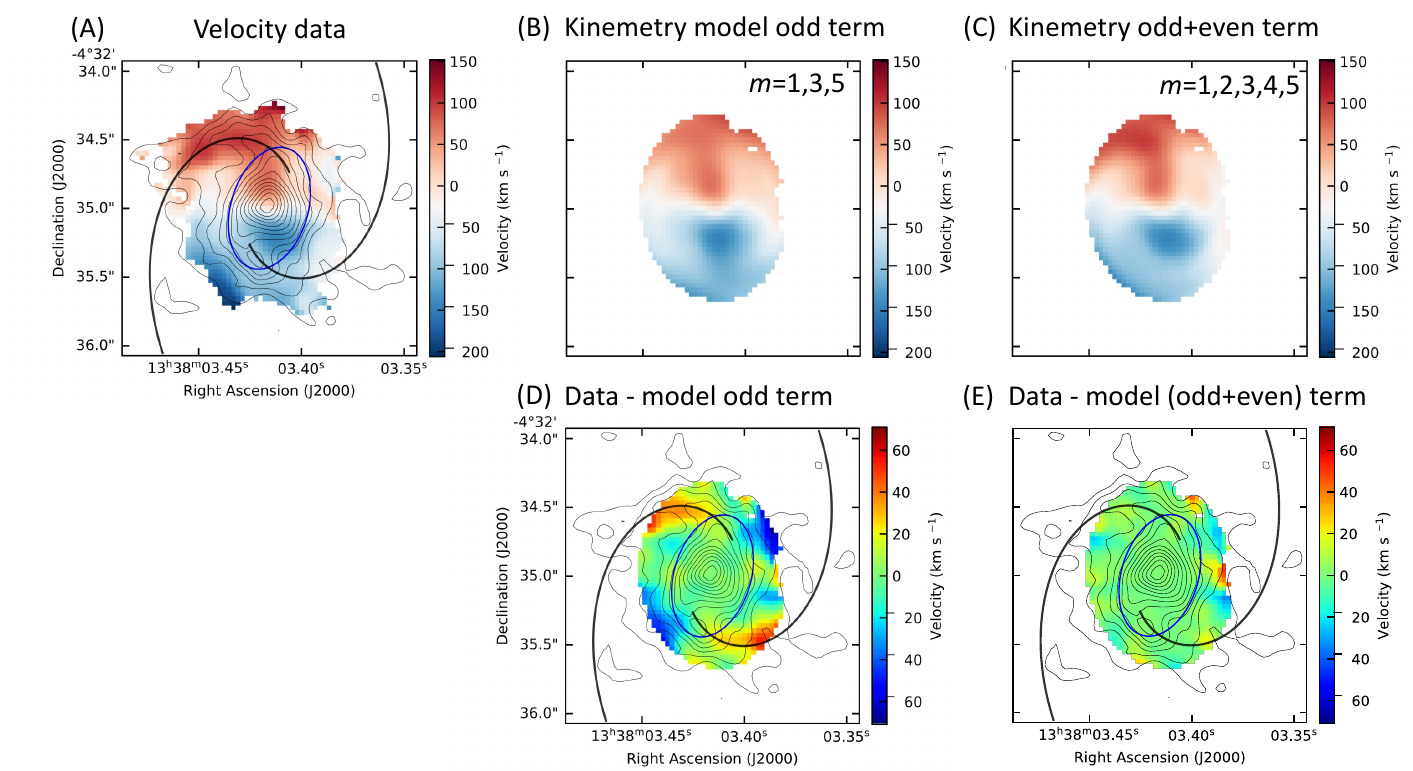}
    \caption{
    \textbf{A}: [C\textsc{ii}] velocity field derived fitting a Gaussian at each spatial pixel of the data cube. \textbf{B}: Best fit \textsc{kinemetry} model using odd Fourier series ($m=1, 3, 5$) to expand the azimuthal velocity profile. 
    \textbf{C}: Best fit \textsc{kinemetry} model using odd + even Fourier series ($m=1, 2, 3, 4, 5$).
    \textbf{D}: Residual map subtracting the model with odd terms only (B) from the data velocity field (A). \textbf{E}: Residual map subtracting the model with both odd and even terms (C) from the data velocity field (A). In \textbf{A}, \textbf{D}, and \textbf{D}, [C\textsc{ii}] contours, bar ellipse, and two armed-spiral are overplotted in the same way as in Fig.~\ref{fig:Fig2}.}
    \label{fig:Fig3}
\end{figure*}

We analyzed the [C\textsc{ii}] gas velocity field of BRI 1335-0417 (Fig.\ref{fig:Fig3}A) using \textsc{kinemetry} \citep{Krajnovic2006-sb}\footnote{See the derivation of [C\textsc{ii}] velocity field \citep{Tsukui2023-wo}, Briefly, we derived the velocity field by fitting a Gaussian function (including Hermite parameters $h3$ and $h4$; \citealt{Cappellari2017-zq}) to the [C\textsc{ii}] spectrum at each pixel, where the 3 channels are available with the emission detected more than 3$\sigma$.} to determine if there is a dynamical imprint from a bar, recent interaction, and or inflow/outflow in the gas kinematics. \textsc{kinemetry} expands the velocity field profile $v(a,\theta)$ along ellipses into Fourier series; 
\begin{equation}
    v(a, \theta)=A_0(a)+\sum_{m=1}^{N} A_m(a)\sin(m\theta)+B_m(a)\cos(m\theta)
\end{equation}
where the ellipses are defined by the semi-major axis $a$, axis ratio $q$, and position angle (PA) from which azimuthal angle $\theta$ is measured. 
As the velocity field of a rotating disk can be expressed by $v(a,\theta)=A_0+B_1(a)\cos(\theta)$ \citep{Krajnovic2006-sb}, the deviation from pure rotational motion manifests as higher order coefficients ($A_1$, $A_2$, $B_2$, ...). By this approach, we can investigate non-circular velocities in a non-parametric manner, in contrast to the dynamical modelling method used in \citet{Tsukui2021-vl}, where they assumed a pure circular rotation and bulge-disk mass profile. 

We performed the \textsc{kinemetry} expansion of the azimuthal velocity profile up to the 5th-order Fourier terms (m=5) along concentric ellipses. We used ellipses with a position angle $7.6^\circ$ and an axis ratio $q=0.79$, which are determined as the global kinematic position angle using \textsc{kinemetry} and the photometric axis ratio of the dust continuum distribution in \citet{Tsukui2021-vl}, consistent with the independent measurements by \citet{Roman-Oliveira2023-uh}. The centre of each ellipse is fixed at the peak position of the continuum image, which matches with the centre of rotation \citep[see Fig.~\ref{fig:Fig3}A][]{Tsukui2021-vl} and the optical quasar position \citep{Tsukui2023-wo}. 

Fig.~\ref{fig:Fig3}B shows the best-fit expansion with odd-term harmonics $m=1, 3, 5$. The data-model residual (Fig.~\ref{fig:Fig3}D) reveals the significant even ($m=2$) component, with two redshifted regions and two blueshifted regions in symmetric positions. The redshifted residuals spatially coincide with the two-armed spiral morphology in the [C\textsc{ii}] and FIR continuum maps. The features are already visible in the velocity field before subtraction in Fig.~\ref{fig:Fig3}A; the redshifted/blueshifted high velocity/low velocity are aligned with spiral arms. 

By definition, the $m=2$ mode cannot be compensated by any odd modes which are the orthogonal basis. This is illustrated by Fig.~\ref{fig:Figb1}~to~\ref{fig:Figb3}. Fig.\ref{fig:Figb1} shows the ellipses along which the azimuthal velocity profiles are extracted by \textsc{kinemetry}. Fig.~\ref{fig:Figb2} and Fig.~\ref{fig:Figb3} show (top) the azimuthal velocity profiles $v(\theta)$ for ellipses at radii 0.54 and 0.68 arcsec with the best fit circular velocity $A_0+B_1\cos{\theta}$ and the circular velocity residual $v(\theta)-A_0+B_1\cos{\theta}$ with higher order expansion with only odds terms and odds plus even terms (bottom), clearly showing that $m=2$ mode is required to reproduce the data profile.

We also confirm that the $m=2$ mode, with the amplitude of up to $\sim$ 30-40 km s$^{-1}$, cannot be diminished by changing the position angle and axis ratio of the kinematic ellipses by $\pm20^\circ$ and $\pm0.1$, respectively (Fig.~\ref{fig:Figb4}). The addition of the even components fully characterizes the velocity field of the BRI 1335-0417 (see the best-fit harmonic expansion in Fig.~\ref{fig:Fig3}C) leaving only a small data-model residual over the disk shown in Fig.~\ref{fig:Fig3}E. Recently, \citet{2023arXiv231110268B} also explored the importance of the even components in spatially resolved gas kinematics for galaxies at intermediate redshift $z\sim0.3$.

\subsection{Interpretation of the \textit{m}~=~2 mode in velocity field}
\add{We conclude that the observed m=2 kinematic mode in the velocity residual is less likely due to inflow or outflow, as symmetric flow would show line of sight velocity residuals \addr{with the opposite sign in symmetric positions \citep{Di_Teodoro2021-ll, Genzel2023-lj}}, leaving the peculiar possibility where the residual on one side is due to outflow while the opposite side is due to inflow. Similarly, it is unlikely due to in-plane motion induced by \addr{the dynamical influence of spiral and bar structures \citep[e.g.,][]{Grand2016-ys, Gomez2021-yx, Monari2016-ro}}. The analytical model suggests a $m$-fold symmetric \add{mass density structure} \addr{would only induce $m-1$ or $m+1$ velocity field distortions \citep{Canzian1993-eb, Schoenmakers1997-gw}\add{, inconsistent} with the observed $m=2$ mode velocity residual for a $m=2$ spiral and bar density structures in the galaxy.} Also, the $m=2$ mode velocity residual cannot be attributed to a PA change of the disk due to disk tilting.} 

\add{Using an analytic model for a galaxy with similar rotational velocity and inclination to BRI 1335-0417, \citet{Gomez2021-yx} show that the radial velocity induced by the spiral density structure is small ($\ll10$ km s$^{-1}$) compared to the disk vertical velocity caused by an interaction. Assuming the line of sight velocity residual of the $m=2$ mode, $\sim$ 35 km s$^{-1}$ (Figure 3D), is purely due to in-plane radial motion, the induced radial velocity would be 57 km s$^{-1}$ after inclination correction. With and without inclination correction the magnitude of the velocity residual seems too large to attribute for the radial motion caused by the spiral structure \citep[e.g.,][]{Grand2016-ys, Monari2016-ro}.}

\add{A} more plausible explanation for the observed $m=2$ mode signature is vertical motion relative to the disk (disk bending mode) induced by external forces, such as recent interactions with satellite galaxies \citep[e.g.,][]{Gomez2017-ug, Bland-Hawthorn2021-fh} \add{or misaligned gas accretions \citep{Khachaturyants2022-ip}}. \add{Strikingly, the kinematic $m=2$ mode in BRI 1335-0417 spatially coincides with the spiral structure in \add{the} intensity map, as shown in Fig. \ref{fig:Fig3}D}. \add{This overlap} is consistent with the vertical wave of stellar and gas disks induced by satellite interaction\add{s in some simulations} \citep{Tepper-Garcia2022-sh}. The simulations reveal the $m=1$ mode in the disk vertical velocity develops $\sim100$~Myr after the Milky-Way-like galaxy is perturbed by the high-speed encounter of the Sagittarius dwarf system, followed by the development of the $m=2$ mode vertical velocity \add{as soon as 200-400~Myr} \citep{Bland-Hawthorn2021-fh, Tepper-Garcia2022-sh}. The vertical velocity pattern initially aligns with the spiral pattern in density and then decouples.

%This is supported by N-body/hydrodynamical simulations \citep{Tepper-Garcia2022-sh} that study the vertical wave of the stellar and gas disks induced by the satellite interaction. The simulations reveal the $m=1$ mode in the disk vertical velocity develops $\sim100$~Myr after the Milky-Way-like galaxy is perturbed by the high-speed encounter of the Sagittarius dwarf system, followed by the development of the $m=2$ mode vertical velocity around 400~Myr. The pattern in the vertical velocity resembled spiral arms. The vertical velocity pattern initially aligns with the spiral pattern in density and then decouples. N-body simulation in \citet{Bland-Hawthorn2021-fh} shows the $m=2$ mode develops in the central region of the disk already in $\sim$200~Myr (see Fig.6 bottom-left in \citealt{Bland-Hawthorn2021-fh}).  

%The fact that the velocity residual and intensity both exhibit a spiral arm pattern in the same location suggests that there is a shared reason for this occurrence

The \add{observed} co-spatial spiral arm patterns in velocity residual and intensity \add{(Fig. \ref{fig:Fig3}D)} are consistent with the simulation \add{of bending waves induced by interactions, or} at least imply a common cause for the two. \add{Given that BRI 1335-0417 has a gas-rich, dynamically hot disk with velocity dispersion of $\sim70$ km s$^{-1}$ \citep{Tsukui2021-vl}, conditions typically unfavourable for spontaneous spiral arm formation \citep{Elmegreen2014-xx}, i}t is reasonable to interpret that the external perturbation responsible for the $m=2$ bending mode induced the observed spiral arms \add{even in a dynamically hot disk} \citep{Law2012-kf} and \add{potentially formed} bar \citep{Lokas2014-pd, Lokas2016-nx}. The disk is \add{turbulent but still} baryon-dominated and gravitationally unstable \citep{Tsukui2021-vl}, making it prone to forming such substructures through perturbations \citep{Law2012-kf}.

The gas velocity dispersion, $\sigma$, of BRI 1335-0417 is estimated to be $71^{+14}_{-11}$~km~s$^{-1}$ \citep{Tsukui2021-vl}. This value is relatively high compared to average values at lower redshifts (45~km~s$^{-1}$ at $z=2.3$, 30~km~s$^{-1}$ at $z=0.9$; \citealt{Ubler2019-lq}, \add{but is in agreement with the scatter seen around cosmic noon} \citep{Kassin2012-er, Wisnioski2015-ah, Wisnioski2019-jp, Ubler2019-lq}. Using the EAGLE simulations, \citet{Jimenez2023-ze} highlights the importance of the gas accretion rate and its relative orientation to the disk in driving the redshift evolution of disk velocity dispersion. The gas accretion (and accompanying satellites) with a large angle relative to the disk plane effectively increases the velocity dispersion of the disk. In general bending mode (buckling instability) is suppressed for a disk with isotropic velocity dispersion \add{\citep[][see for review \citealt{Sellwood2013-xn}]{Toomre1964-qf, Araki1985-ga, Merritt1994-nw}} which may hold for high-redshift galaxies \citep{Genzel2023-lj, Genzel2017-xu}. The bending wave induced by the vertical perturbation of the high-angle accreters may be immediately damped and contribute to the kinetic energy of the disk velocity dispersion. 

%The scenario that BRI 1335-0417 has experienced an interaction $\sim$100-200~Myr ago triggering the currently observed high SFR, aligns with the rough estimate of the starburst lifetime, the gas depletion time $\sim$50-200~Myr \citep{Tsukui2023-fv}. The simulated galaxies in \citet{Tepper-Garcia2022-sh, Bland-Hawthorn2021-fh} vastly differ from BRI 1335-0417 in terms of various parameters, such as gas fraction, total mass, gas velocity dispersion, and satellite separation. The time scale of 200-400~Myr to develop the m = 2 mode in the simulation may be shorter $\sim$ 100-200~Myr for this galaxy considering the dynamical time of BRI 1335-0147 is $\sim120$~Myr (orbital period at effective radius), two times smaller than that of the galaxy in the simulation.

\section{Conclusion}
We identify an elongated bar-like structure in the z=4.4 spiral galaxy BRI 1335-0417 using both [C\textsc{ii}] and \add{rest-frame 160$\micron$} far infrared (FIR) images by fitting ellipses to the isophotes. The variation of ellipticity and position angle shows the characteristic profile of barred galaxies. The identified bar, $3.3^{+0.2}_{-0.2}$ kpc long in radius, appears to bridge the two arm spirals identified in the previous study \citep{Tsukui2021-vl}. When compared to the disk scale length, the bar length is larger than stellar bars in nearby galaxies \citep{Erwin2019-im}.

The [C\textsc{ii}] bar is more elongated than the FIR bar and displaced towards the leading edge of the FIR bar (at the downstream side of the gas motion), assuming the spiral arms of BRI 1335-0417 are trailing spiral arms. If we \add{attribute the structural difference of FIR and [C\textsc{ii}] to the fact that FIR emission can depend on the distribution of both dust (interstellar medium roughly traced by [C\textsc{ii}]) and young massive stars heating the dust}, the observations \add{seem to} align with the established picture from both observations and simulations -- the compressed gas flow (traced by [C\textsc{ii}]) forms a straight line into the centre of the galaxy along the leading edge of the \add{young} stellar bar (more traced by FIR emission).

The galaxy is shown to have a significant gas fraction (>73\%), hosting the star-forming gas bar\add{, which prevents us from assuming the presence of a stellar bar dominating the gravitational potential and influences the gas bar kinematics.} 
%seems inconsistent with the simulation result in general; gas content suppresses stellar bar formation out of the stellar disk, alternating the stellar orbit of disk stars.} 
Along with this object, the abundant bar-like structures in FIR continuum observed in bright sub-millimetre galaxies \citep{Hodge2019-uo, Gullberg2019-ec} and another example high-redshift barred galaxy at z=4.3 with an extremely high gas fraction \citep{Smail2023-gm} may prompt a different perspective on the bar formation and dynamics rather than simulations focused on how disk stars form stellar bars. The dominant gas disk\add{, by itself gravitationally unstable,} potentially leads to the formation of a gas bar \citep{Barnes2001-sj}, as an axisymmetric 100\% gas disk is theoretically proven to form a self-gravitating stable gaseous bar structure \citep[][]{Cazes2000-eu}.

By applying a harmonic expansion of azimuthal profiles of the [C\textsc{ii}] velocity field with \textsc{kinemetry} we reveal a dominant $m=2$ mode with the amplitude of up to $\sim 30-40$~km s$^{-1}$ in the velocity \add{residual}. \addr{The $m=2$ mode cannot be explained by large-scale inflow/outflow or the non-circular motions caused by bar/spirals, which generally produce odd-numbered modes \citep[e.g.][]{Canzian1993-eb, Schoenmakers1997-gw}.} \add{Therefore we interpret the $m=2$ mode as vertical motion relative to the disk (disk bending wave).} The redshifted velocity regions contributing to the $m=2$ mode coincide spatially with the spiral arms, consistent with the scenario where the galactic disk is perturbed by an interaction creating the initially co-spatial spiral density wave and vertical bending mode in the disk \citep[e.g.,][]{Bland-Hawthorn2021-fh, Tepper-Garcia2022-sh}. 

As it is assumed that a gas disk is stable for the bending mode (buckling instability), the $m=2$ mode velocity perturbation and spiral arms are likely to be induced by \add{a} recent interaction and/or gas accretion. It is natural to suppose that \add{such an} interaction \add{would also} activat\add{e} the high star formation activity. 

For the first time, this study detects an unambiguous $m=2$ mode in a disk at high redshift ($z>4$), lending support to the use of disk seismic ripple as evidence of a recent strong perturbation. With further numerical simulations tuned for galaxies with realistic gas, halo, and stellar masses and gas accretion histories, the $m=2$ mode may provide a useful constraint on the exact timing and primary origin of external perturbations.

\section*{Acknowledgements}
TT is grateful to the conference organizers of Galactic Bars: driving and decoding galaxy evolution held in Granada, Spain in July 2023, which greatly helped in writing this paper and Trevor Mendel for encouraging me to attend. TT also thanks Camila de S\'{a}-Freitas, Karin Menedez Delmestre, Ewa Luiza Lokas, and Andreas Burkert for insightful discussions. This research was supported by the Australian Research Council Centre of Excellence for All Sky Astrophysics in 3 Dimensions (ASTRO 3D), through project number CE170100013. Data analysis was partly carried out on the Multi-wavelength Data Analysis System operated by the Astronomy Data Center (ADC), National Astronomical Observatory of Japan. This paper makes use of the following ALMA data: ADS/JAO.ALMA\#2017.1.00394.S. ALMA is a partnership of ESO (representing its member states), NSF (USA) and NINS (Japan), together with NRC (Canada), NSC and ASIAA (Taiwan) and KASI (Republic of Korea), in cooperation with the Republic of Chile. The Joint ALMA Observatory is operated by ESO, AUI/NRAO and NAOJ. Finally, TT thanks Takaho Masai for his kind support at NAOJ.

%%%%%%%%%%%%%%%%%%%%%%%%%%%%%%%%%%%%%%%%%%%%%%%%%%
\section*{Data Availability}
The ALMA data we use in this work are publicly available in \url{https://almascience.nrao.edu/aq/}. 



\bibliographystyle{mnras}
\bibliography{example} % if your bibtex file is called example.bib

%\begin{thebibliography}{99}
%\end{thebibliography}

%%%%%%%%%%%%%%%%%%%%%%%%%%%%%%%%%%%%%%%%%%%%%%%%%%

%%%%%%%%%%%%%%%%% APPENDICES %%%%%%%%%%%%%%%%%%%%%

\appendix

%\begin{figure}
%	\includegraphics[width=\columnwidth]{Fig1_isophotes.pdf}
%    \caption{[C\textsc{ii}] and FIR images of BRI 1335-0417 overlaid with the best-fitting ellipses in black, the bar ellipse and its PA in the solid blue line, and the disk in the dashed blue line.}
%    \label{fig:Fig1a}
%\end{figure}

\section{Supplemental information of \textsc{kinemetry} analysis}
\begin{figure}
	\includegraphics[width=\columnwidth]{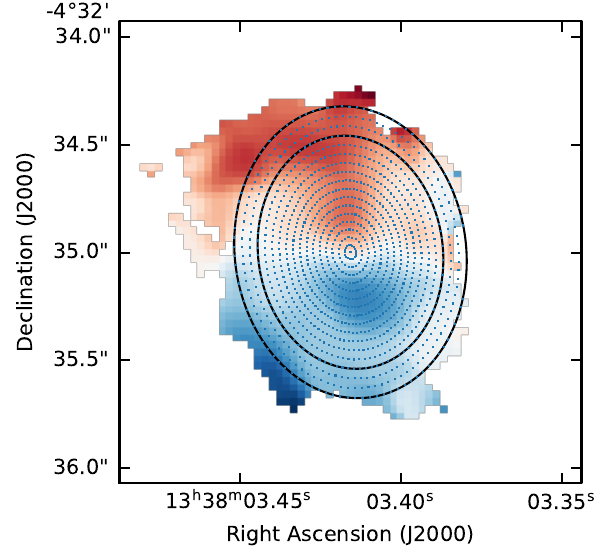}
    \caption{Blue points: Ellipses and azimuthal data sampling of each ellipse used for \textsc{kinemetry} analysis overlain on the BRI 1335-0417 velocity field. Black solid lines show ellipses with a radius of 0.54 arcsec and 0.68 arcsec whose velocity profiles are shown in \ref{fig:Figb2} and \ref{fig:Figb3}.}
    \label{fig:Figb1}
\end{figure}

\begin{figure}\includegraphics[width=\columnwidth]{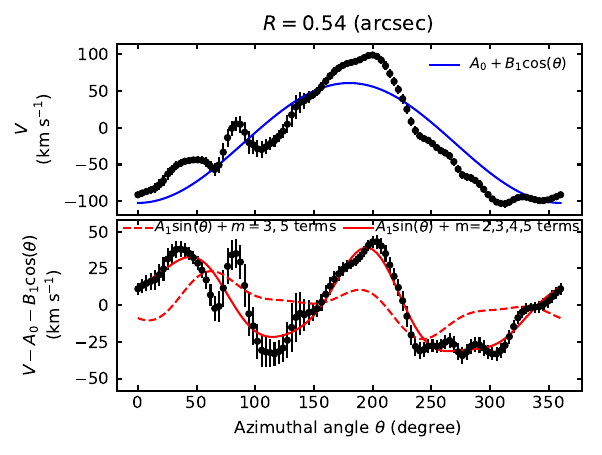}
    \caption{Top: Azimuthal velocity profile $v(\theta)$ at a radius of 0.54 arcsec (black points) with the best fit circular velocity $A_0+B_1\cos{\theta}$ (blue). Bottom: the residual velocity after subtracting the circular velocity $v(\theta)-A_0+B_1\cos{\theta}$ (black points) and the best-fit expansion with higher order harmonics. The red dashed line is the best fit with only odd terms ($m=1,3,5$) and the solid line is the best fit including even terms ($m=1,2,3,4,5$).}
    \label{fig:Figb2}
\end{figure}

\begin{figure}\includegraphics[width=\columnwidth]{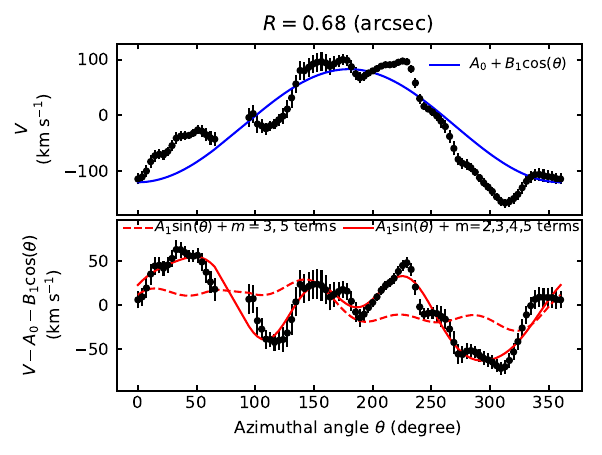}
    \caption{Same as \ref{fig:Figb2} but for a radius of 0.68 arcsec.}
    \label{fig:Figb3}
\end{figure}

\begin{figure}\includegraphics[width=\columnwidth]{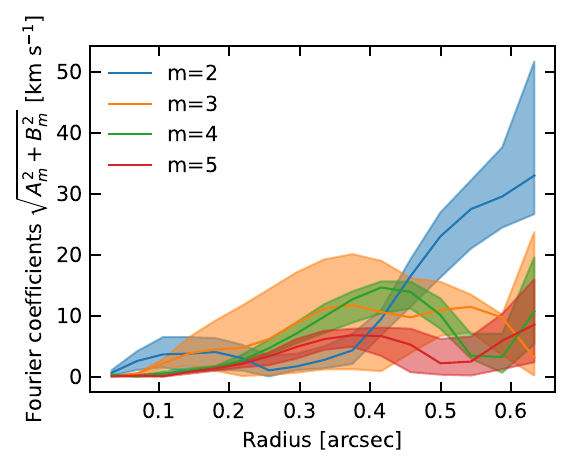}
    \caption{Solid lines; Fourier coefficients (m=2, 3, 4, 5) of azimuthal velocity profile as a function of the semi-major axis of the ellipse obtained by \textsc{Kinemetry}, using concentric ellipse with an axial ratio q=0.79 and a position angle of 7.6$^\circ$. Shaded regions; the maximal variation allowed when varying the axial ratio ($\pm 0.1$) and position angle ($\pm 20^\circ$) of the ellipse used sampling the azimuthal velocity profile.}
    \label{fig:Figb4}
\end{figure}

In this section, we provide the detailed results of the \textsc{kinemetry} analysis to verify the presence of $m=2$ mode. Fig.~\ref{fig:Figb1} shows the ellipses to extract the azimuthal profile of the velocity field. Fig.~\ref{fig:Figb2} and Fig.~\ref{fig:Figb3} illustrate the harmonic expansion of the azimuthal velocity profiles at two specific radii: 0.54 arcsec and 0.68 arcsec. For the sake of visualization, the pure rotational motion (top: blue line) is subtracted from the original velocity profile (top: black points) and we have shown a higher-order expansion for the residual (bottom: black points). The expansion involves odds terms with or without even terms up to $m=5$ modes (shown as red solid and dashed lines respectively). Even terms, especially m=2 mode are required to reproduce the velocity fields. The dominance of the $m=2$ relative to the other modes obtained by expansion with full terms (odds plus even) is shown in Fig.~\ref{fig:Figb4}; the m=2 is dominant after 0.4 arcsec and, opposed to other odds modes, is not diminished by the vast change of PA and ellipticity of sampling ellipses.  

%%%%%%%%%%%%%%%%%%%%%%%%%%%%%%%%%%%%%%%%%%%%%%%%%%

% Don't change these lines
\bsp	% typesetting comment
\label{lastpage}
\end{document}